\documentclass[article,aps,prb]{revtex4}
\usepackage{epsf,graphicx}
\begin{document}
\title{{\bf Chirality and Protein Folding}}

\author{Joanna I. Kwieci{\'n}ska and
Marek Cieplak\footnote{
Correspondence to: Marek Cieplak,
Institute of Physics, Polish Academy of Sciences,
Al. Lotnik\'ow 32/46, 02-668 Warsaw, Poland;
Tel:  48-22-843-7001, Fax:  48-22-843-0926;
E-mail: {\sl mc@ifpan.edu.pl}}
}
\affiliation{ Institute of Physics, Polish Academy of Sciences,
Al. Lotnik\'ow 32/46, 02-668 Warsaw, Poland}

\begin{abstract}
There are several simple criteria of folding to a native state in 
model proteins. One of them involves crossing
of a threshold value of the RMSD distance away from the
native state. Another checks whether all native contacts are
established, i.e. whether the interacting amino
acids come closer than some characteristic distance.
We use Go-like
models of proteins and show that such simple criteria
may prompt one to declare folding even though fragments
of the resulting conformations have a wrong sense of chirality.
We propose that a better condition of folding should augment
the simple criteria with the requirement that most
of the local values of
the chirality should be nearly native.
The kinetic discrepancy between the simple and compound criteria
can be substantially reduced in the Go-like models by providing
the Hamiltonian with a term which favors native values of the
local chirality. We study the effects of this
term as a function of its amplitude and compare it to other models
such as with the side groups and with the angle-dependent potentials.
\end{abstract}

\maketitle





\section{Introduction} \label{int}

Theoretical studies of the kinetics of protein folding in off-lattice
models encounter a difficulty when it comes to making a decision
about at what stage of the
temporal evolution from an unfolded state the system can be
considered as having reached the native conformation.
The native conformation is usually known within some experimental
resolution but, in models, this is a state of "zero measure"
in the three dimensional space.
Criteria of folding must then involve some
finite effective spatial extension of the native conformation
and then checking whether an evolving conformation 
has already entered the resulting native "cocoon".
An example of providing a way to define such an cocoon is
given in \cite{Li1999,Li1999a,Hoang2000,Hoang2001}
where one studies features in time evolution of
distortions in shape of the native conformation. The
shape distortion method, however,
turned out to be rather subtle and applicable, in practice,
only to secondary structures of proteins.\\

Simple and widely used criteria of folding postulate existence of 
a variable that crosses a predefined threshold value on folding.
One example of such a variable is the RMSD distance away from
the native state. Another is the fraction, $Q$, of the established
native contacts which implies descending below some selected value
of the total energy.
Usually, the threshold value of $Q$
is selected to be equal to 1. Note, however,
that the very notion of an established
contact itself depends on two amino acids coming closer together
than some threshold value, $r_c$, of their mutual distance.
In the following, the folding criterion based on the RMSD distance
will be denoted by R and the criterion based on contacts by Q.
For completeness, we also consider a third criterion, denoted
by A, which is based on the angular properties of the protein
backbone. Criterion A involves checking whether the bond and
dihedral angles are at their native values within a chosen range.\\

In this paper, we use coarse-grained Go-like models of proteins
\cite{Abe1981,Takada1999} with the Lennard-Jones potentials in the
contacts. We show that all three of these simple 
criteria may prompt one to declare folding even though fragments
of the resulting conformations have a non-native sense of chirality --
a point that has already been noted by Ortiz, Kolinski, and Skolnick
\cite{Kolinski}.
Figure 1, for crambin, shows an example of this problem when the
Q criterion is used.
It is seen that the folding process may generate a left handed 
$\alpha$-helix instead of the right-handed one that is actually
present in the true native conformation.
This misfolding event occurs despite
the fact that all native contacts are found to be within
their respective values of $r_c$ (1.5 times the characteristic
length parameter $\sigma$ in the potential) 
and the RMSD distance away from the native state
is only around 2.5 $\AA$.
In order to counter this difficulty with the
folding criteria one may consider
adopting more stringent values of the threshold parameters but this
approach turns out to be impractical for it usually leads to prohibitively
long time evolution, i.e. it results in the lack of folding.
Even though our conclusions are drawn from 
studies of Go-like models, we expect that similar problems
are also encountered in all-atom simulations.\\

We propose that a reasonable condition of folding should combine
the simple criteria with the requirement that all local values of
the chirality should be nearly native. 
We show that setting of the local chiralities in native-like values
may or may not precede the "calls" 
resulting from the simple criteria, depending on the criterion and
the type of a protein, but the compound requirement allows for
more accurate studies of folding and typically
generates structures that appear to be of correct topology.\\

The kinetic discrepancy between the simple and compound criteria
can be substantially reduced in the Go-like models by providing
the Hamiltonian with an extra term which favors native values of the
local chirality as introduced in \cite{Cieplak2003a}.
We study the effects of this
term as a function of its amplitude and select its value that could
be used in simulations.\\


As will be made explicit later, the definition of the local chirality 
involves coordinates of four subsequent $C^{\alpha}$ atoms.
It should be noted that a definition of the bond angle
involves three $C^{\alpha}$ atoms and that of the dihedral
angle -- four. 
One may ask then two questions: First, how does the chirality based criterion
relate to the criterion involving the bond and dihedral angles?
Second, is it better to use potentials that favor the native
sense of chirality or those which favor the native values of the
bond and dihedral angles?
We discuss these issues at the end of the paper and find the
chirality potential to act in a comparable way to the
angular potential. However, its usage is much more economical
in simulations. We also find that
simple modelling of the side groups by the $C^{\beta}$ atoms
does not automatically take care of making the chirality
in the folded state to be correct.\\

\vspace*{0.5cm}
\section{The Go-like model}

We perform molecular dynamics simulations of a continuum space
coarse-grained Go-like model. The Go-like models are defined
through experimentally determined native conformations of proteins
as stored in the Protein Data Bank \cite{Bernstein1997}.
The details of our approach are described in 
Ref. \cite{Hoang2000,Hoang2001}  
Each amino acid is represented by a point particle of mass $m$ located
at the position of the C$^{\alpha}$ atom.
The interactions  between amino acids are divided into native and 
non-native contacts. 
We determine the native contacts by considering the
all-atom native structure and by identifying those pairs
of amino acids in which there is an overlap between effective
spheres that are associated with heavy 
atoms \cite{Cieplak2002,Cieplak2003a}. 
These spheres have radii
that are a factor of 1.24 larger
than the atomic van der Waals radii 
\cite{Tsai1999}  to account for the softness
of the potential. The native contacts are then represented
by the Lennard-Jones potentials
$4 \epsilon [ (\sigma_{ij}/r_{ij})^{12} - (\sigma_{ij}/r_{ij})^{6}]$,
where $r_{ij}\;=\; |\vec{r}_i -\vec{r}_j|$ is the distance between 
C$^\alpha$ atoms $i$ and $j$ located at $\vec{r}_i$ and $\vec{r}_j$ 
respectively.
The length parameters $\sigma _{ij}$
are determined so that the minimum of the pair potential coincides
with the distance between C$^\alpha$ atoms in the native structure.
In order to prevent entanglements, the remaining pair-wise
interactions, i.e. the non-native contacts, correspond to a
pure repulsion. This is accomplished by taking the Lennard-Jones
potential with $\sigma_{ij}=\sigma=5$ $\AA$ and truncating it
at $2^{1/6}\sigma$.\\

All contacts have the same energy scale $\epsilon$. This energy
scale corresponds to between 800 and 2300 K as it effectively
represents hydrogen bond and hydrophobic interactions. The room
temperature should then correspond to $\tilde{T}=k_BT/\epsilon$ of
about 0.1 -- 0.3 ($k_B$ is the Boltzmann constant). 
Neighboring C$^\alpha$ atoms are tethered by a harmonic potential
with a minimum at 3.8 $\AA$ and the force constant of
$ 100 \epsilon $\AA$^{-2}$.\\

The equations of motion are integrated using a fifth-order predictor-corrector
algorithm with time step $dt = 0.005 \tau$.
A Langevin thermostat with damping constant $\gamma $ is coupled
to each C$^\alpha$ to control the temperature.
For the results presented below $\gamma = 2 m/\tau$, where
$\tau =\sqrt{m \sigma^2 / \epsilon} \sim 3$ps is the characteristic time
for the Lennard-Jones potential.
This produces the overdamped dynamics
appropriate for proteins in a solvent,\cite{Cieplak2003a}
but is roughly 25 times smaller than the realistic damping from water,
\cite{Veitshans1997}.
Tests with larger $\gamma$ confirm a linear
scaling of folding times with $\gamma$.
\cite{Hoang2000,Hoang2001}  so
the folding times reported below should be
multiplied by 25 for a more meaningful comparison to experiment.\\

The folding processes are characterized by the order in
which native contacts are formed.
At a finite $T$, a pair distance $r_{ij}$
may fluctuate around a selected cut-off value. Thus, when
discussing folding using the contact criterion, we determine
the average time $t_c$ for each contact to form for
the first time.  Unless stated otherwise, the cut-off
value for the presence of a contact between
amino acids $i$ and $j$ we take 
1.5$\sigma _{ij}$ in model proteins and 1.36$\sigma _{ij}$
in model secondary structures. The latter value is equal to the
inflection distance for the Lennard-Jones potential. This choice 
of the prefactor does
not matter much for the secondary structures but choosing the inflection
distance for proteins makes the folding process prohibitively long
lasting. Note that the cut-off $r_c$ depends on the pair of amino acids
in our model.\\

The chirality of residue $i$ is defined as
\begin{equation}
C_{i}=\frac{(\vec{v}_{i-1}\times \vec{v}_{i})
\cdot \vec{v}_{i+1}}{d_{0}^{3}} \;\;, 
\label{ci}
\end{equation}   
where $\vec{v}_{i}=\vec{r}_{i+1}-\vec{r}_{i}$, and $d_{0}=|v_{i}|$ 
is a distance between 
neighbouring residues as represented by the $C^{\alpha}$ atoms.
For a protein of $N$ amino acids, $C_i$ is defined for $i$ from 2 to
$N-2$. In Equation (1), the amino acids are labelled from the terminal
N to the terminal C. It is possible to consider an alternative
definition which would involve proceeding in the opposite way
and which would define chiralities for $i$ from 3 to $N-1$. Another
possibility is to consider a symmetrized combination of the two
definitions. We have found, however, that these alternative variants
work quite similar to the basic definition.\\

In the unlikely event in which all
all of the atoms involved in the determination of $C_i$ are located in
a plane one gets a $C_i$ which is equal to zero. 
Otherwise $C_i$ is positive or
negative for the right-handed and left-handed local turns
respectively. The magnitudes are small in nearly planar loop regions.
The distribution of $C_i$ in proteins is essentially
bimodal \cite{Cieplak2003a}. A maximum around 0.7
corresponds to $\alpha$-helices. We will show that, in $\beta$-sheets
$C_i$ ranges between -0.15 and 0.15. In turns and loops, $C_i$
can also be either positive or negative. \\

We illustrate the findings of our studies by focusing on
the (16-31) helical fragment of the P chain of the capsid protein P24 --
1e6j, the (41-56) hairpin of the protein 1pga and on crambin -- 1crn.
The folding times were determined as the median first passage time.
501 different trajectories were used in the case
of secondary structures  and at least 101 trajectories
in the case of crambin.\\

\section{Criteria of folding} 

Consider 1e6j(16-31) -- the helical fragment of 16 monomers.
Figure 2 shows examples of conformations that
are declared folded according the criteria Q, R, and A. The first
two criteria were specified in the Introduction. As to the A criterion,
we require that the local bond, $\theta _i$, and dihedral, $\phi _i$,
angles are within $\pm \Delta \theta$ and $\pm \Delta \phi $ away
from the native values respectively. We take rather generous
$\Delta \theta$ and $\Delta \phi$ of 60$^o$ 
in order to get folding times
which are comparable to those obtained through the other two criteria.
Values which are substantially more stringent usually yield no 
folding in computationally accessible time.\\

It is seen that each of the dynamically obtained example conformations
has regions, usually at the terminals, which twist in the opposite sense
relative to what is found in the native conformation. This 
misfolding phenomenon is captured by the values of the local
chiralities that are listed in the caption of Figure 2.\\

One way to characterize this phenomenon is to count the number, $b$,
of $C_i$'s that have a sign which is opposite to the native sign. The
distributions of $b$ for the helix and for the (41-56) hairpin fragment 
of the protein 1pga are shown in the top panels of Figure 3. 
The zero value of $b$ means the absence of chirality defects. It is 
seen that the distributions depend on the folding criterion used 
and A is found to be the most successfull
in this respect: $b$ is concentrated 
at values not exceeding 3. The Q and R criteria
work in a comparable way for the hairpin, but Q is worse for the
helix  (for several other helices it was found to be comparable).
The bottom panels of Figure 3 illustrate the sequantial
distribution of the local chirality defects, $b_i$. ($i$ is counted
here from the begining of the fragment and not from the N-terminal
of the host protein).
In the helix, the three criteria generate defects throughout the
chain, though, for this
particular helix, A favors defects arising closer to the C-terminal.
On the other hand, the defects generated
in the hairpin as a result of adopting the A criterion are localized
at the center (and do not depend on the example of a hairpin).
The other two criteria favor no particular region  
in the hairpin. \\

In order to monitor the chirality defects dynamically we introduce
the parameter $K$ which compares the values of $C_i$ to the
native values, $C_i^{NAT}$, and counts the values which 
can be considered as being
substantially native-like, i.e. which are of the right sign 
and their magnitudes are at least 50\% of the native strength.
A convenient definition of $K$ is then given by
\begin{equation}
K\;=\;\sum_{i=2}^{N-2} \; \Theta (C_i/C_i^{NAT} \;-\;0.5) \;\;,
\end{equation}
where $\Theta(x)$ is the step function ($\Theta(x)$ is 1 if $x \ge 0$
and 0 otherwise).\\

The top-left panel of Figure 4 shows that the
local values of $C_i$ may have a rapid temporal evolution which translates
into a noisy behavior of $K$. It is thus not realistic to seek the
fully native value $K_{NAT}$ of $N-3$. However, $K_c=0.75\; K_{NAT}$ is
attainable in simulations, as illustrated in the remaining panels of
Figure 4. These panels show examples of the time evolution of the
number of established native contacts, $NQ$, and of $K$ for two 
secondary structures and for crambin. In the case of the helix,
$K$ strikes $K_c$ for the first time in the example trajectory
before $Q$ hits 1 for the first time. The opposite takes
place in the
illustrative trajectories for crambin and for the hairpin.\\

We propose that the shape sensitive criterion of folding should 
combine the simple criteria such as Q, R, and A with a condition
on the chirality:
\begin{equation}
K\;\ge \; K_c \;\;.
\end{equation}
From now on, we focus on the contact based criterion, Q. The compound
contact-chirality criterion will be denoted by Q$_K$. The folding time
in the Q$_K$ criterion is then defined as the first instant when
both $Q=1$ and $K \ge K_c$. Similar conclusions are expected to hold
for the R and A criteria. The Q criterion and its extension
seems to us to be the simplest to use. When employing the R criterion
an appropriate choice of the cutoff RMSD value ought to
involve scaling with the system size as pointed out by
Betancourt and Skolnick \cite{skolnik}. We also observe that conditions
on the angles are especially hard to satisfy in loop regions. \\

The compound criterion Q$_K$ leads, of course, 
to longer folding times than the simple criterion Q.
This is illustrated in Figure 5 which shows
three characteristic times, $t_Q$, $t_K$, and $t_{QK}$ for two
secondary structures and for crambin. The first of these times corresponds
to the median folding time obtained by using the Q criterion.
The second time denotes the median time for $K$ to reach $K_c$
for the first time. Finally, $t_{QK}$ is the folding time obtained
through the compound criterion Q$_K$. Both $t_Q$ and $t_{QK}$
exhibit the ubiquitous U-shaped dependence on the temperature
indicating the presence of the optimal temperature, $T_{min}$,
at which folding proceeds the fastest. The value of $T_{min}$ itself
depends somewhat on the choice of the folding criterion.\\

At the vicinity of the optimal temperature, the establishment of
native-like chirality and of the native-like number of contacts,
separately, is seen to be almost simultaneous for the helix.
Establishing both of these quantities simultaneously takes twice as long.
In the hairpin, the contacts get established before the chirality.
In crambin, there is a clear separation between various stages and the
contacts get established the first.\\

The Q$_K$ criterion leads to a significant reduction in the extension
of the wrong chirality defects compared to the Q criterion.
This feature is llustrated in Figure 6 for the helix and the hairpin.
In the helix, there are no defects with size $b$ of two or larger
(i.e. the left-handed conformations do not arise) and, if present,
they are concentrated at the C terminal where chiralities at the last two
sites are not defined. With the simple Q criterion (Figure 3) the
defects arise at any site and can extend throughout the helix.
The Q$_K$ criterion also introduces a significant improvement
in the conformations found for the hairpin.\\

\vspace*{0.5cm}

\section{The chirality potential}
\vspace*{0.5cm}

The Go-like modelling works with a potential which favors the native 
conformation. In this spirit, it seems sensible to improve on the
current modelling by adding a potential which favors the native sense
of the local chirality.  It should also be noted that atomic level
interactions in proteins do favor specific senses of chirality
that are found in nature.
Following reference \cite{Cieplak2003a}
(there is a misprint in the definition shown there)
we consider the term
\begin{equation}
V^{CHIR}=\sum_{i=2}^{N-2}\frac{1}{2}\kappa \epsilon 
C_{i}^{2}\Theta(-C_i C_{i}^{NAT})\;\;,
\label{potchiral}
\end{equation}
where the dimensionless
parameter $\kappa$ controls the strength of the potential and 
its value needs to be selected. 
The potential $V^{CHIR}$ involves a harmonic
cost in $C_i$ if the local chirality is non-native. We have also
considered an even more Go-like version of $V^{CHIR}$ in which
instead of the step function, there is a harmonic penalty for deviations
from the native values of the local chiralities:
\begin{equation}
V^{CHIR}_1=\sum_{i=2}^{N-2}\frac{1}{2}\mu \epsilon 
(C_{i}\; - \; C_{i}^{NAT})^2\;\;,
\label{potchiral}
\end{equation}
where $\mu$ is a strength parameter.
We have found that either choice of the potential leads
to qualitatively similar results except that $V^{CHIR}_1$
tends to yield broader regions of temperature in which
folding is optimal. The results shown in this chapter are based
on equation (4).\\

Still another way to introduce a chirality-related potential
has been recently proposed by Chen, Zhang, and Ding \cite{Chen}
in their studies of homopolypeptides on a lattice. Their definition
applies to helices and it involves a linear energy cost (i.e.
the chirality acts like an external uniform field and not like
a restoring potential). It may be worthwhile to compare the
molecular dynamics of such a model with our results.\\

When one considers not coarse-grained but atomic models
then the preference for a native chirality can also
be implemented through conditions
of consistency with the Ramachandran maps
\cite{Kolinski1,Kolinski2} (in this reference the maps are
translated into a simplified description of the conformational
space).\\

For a system with the added potential $V^{CHIR}$, the folding times
obtained with the criterion Q$_K$ depend on the value of $\kappa$.
Figure 7, for the hairpin, shows that the dependence saturates
around $\kappa$=1. Furthermore, for $\kappa \ge 1$ the difference
between the folding times obtained with the Q and Q$_K$ criteria 
becomes small. In other words, one can revert to the simple criterion Q
by introducing $V^{CHIR}$ with a sufficiently large $\kappa$.
From now on, we stay with the choice of $\kappa=1$ as used in references
\cite{Cieplak2003a} and \cite{Cieplak2004}. The kinetic equivalence
of Q$_K$ to Q combined with $V^{CHIR}$ is qualitative and it appears to
be valid in the vicinity of $T_{min}$ as shown in Figure 8. Away from
$T_{min}$, $t_Q$ and $t_{QK}$ diverge from each other in the case of the
hairpin. For crambin and the helix, there is no divergence.
It should be noted, however, that the folding times in systems without
$V^{CHIR}$ are longer compared to the case of $\kappa$=0, independent
of the folding criterion used.\\

The contact criterion Q with the chiral potential naturally leads
to small RMSD values at folding as illustrated in Figure 9 for crambin.
This figure also shows that the distribution of the RMSD values on 
folding with the chiral potential and criterion Q is nearly the
same as without the chirality potential but with criterion Q$_K$.\\

One may qualitatively asses the role of the chirality potential
on the folding funnel by plotting the free energy as a function of
$Q$, the fraction of the established native contacts, and $R$,
the binned values of the RMSD. This $F(Q,R)$ for the helix is
shown in Figure 10 at the
temperature of optimal folding.
It is seen that, in the absence of $V^{CHIR}$, the free energy
is endowed with two minima whereas for $\kappa$=1 there is
just a simple folding funnel. This indicates that the
chirality term leads to a smoother shape of the free energy landscape.\\

We now examine the folding process in more details by looking at
establishment of specific contacts, i.e. by studying the so called
folding scenarios. A contact is said to be established if its amino
acids approach each other sufficiently close  -- for the first time.
A chirality in the contact
will be said to be established if both of the amino acids involved
acquire native-like chirality (50\% of the native value of $C_i$).\\

The folding scenarios of the helix, hairpin and crambin are shown in 
Figures 11 through 13 where the case of $\kappa$=1 is compared to 
that of $\kappa$=0.
In the case of the helix with $\kappa$=1, 
the average times to establish
pair-wise chirality are nearly site independent and practically
equal to times needed to establish the contacts (independent of
the establishment of the chirality). The combined criterion
essentially doubles the times and leads to somewhat shorter
times at the terminals. For $\kappa$=0, there is a stronger
preference for initiation at the C terminal.
In the case of the hairpin, there is a tendency to fold the
structure from the turn outward, independent of the
value of $\kappa$, except that it appears to be difficult
to establish the right chirlaity right at the turn ($i$=7).
The folding scenario of crambin is nearly monotonic as a 
function of the contact order. When $\kappa =1$, the folding events
as measured by the contact establishment nearly coincide with those
determined by the contact-chirality criterion. Thus, in practice,
when the chirality potential is present, there is no need
for the compound criterion as evidenced also by the distribution
of the resulting RMSD (Figure 9).\\

Finally, we comment on the values of the characteristic 
temperatures $T_{min}$ and $T_f$. The latter is the folding temperature
at which the probability of staying in the native 'cocoon'
crosses $\frac{1}{2}$. $T_f$ is a measure of the thermodynamic
stability and its determination involves a definition of the
cocoon, i.e. on whether we use the Q or Q$_K$ criterion to define 
the effective native conformation. The left panel of Figure 14
shows that both characteristic temperatures saturate as a 
function of $\kappa$ and the results obtained with the two criteria
merge. For crambin (Figure 15) and the hairpin (not shown) there
is no merger but there is a saturation beyond $\kappa$=1.\\

\vspace*{0.5cm}

\section{The angular potential}
\vspace*{0.5cm}

We now consider the angle-dependent Go-like potential,
$V^{ANG}\;=\; \Upsilon (V^{BA} + V^{DA}$), where
\begin{equation}
V^{BA}=\sum_{i=1}^{N-2}K_{\theta}(\theta_{i}-\theta_{0i})^{2} \;\;,
\end{equation}
and
\begin{equation}
V^{DA}=\sum_{i=1}^{N-3}[K_{\phi}^{1}(1+cos(\phi_{i}-
\phi_{0i})+K_{\phi}^{3}(1+cos3(\phi_{i}-\phi_{0i}))]\;\;.
\end{equation}
Here,
$\theta_{i}$ and $\phi_{i}$ denote the bond and dihedral angles
respectively and the subscript 0 indicates the native values
that this potential favors.
Following  Clementi, Nymeyer, and Onuchic
\cite{Clementi} (see also a further discussion in
reference \cite{Cieplak2003b}) we take 20$\epsilon$,
$\epsilon$, 0.5$\epsilon$ for $K_{\theta}$,
$K_{\phi}^{1}$ and $K_{\phi}^{3}$ respectively.
The quantity $\Upsilon$ is an overall controll parameter
of the potential strength such that when $\Upsilon$=1 the
customarily used strength is obtained \cite{Clementi}.\\

The determination of the dihedral angles involves four sites
which suggests a formal similarity to $V^{CHIR}$. We now demonstrate
that the effects of the two potentials are different. One example
of the difference is shown in the right hand panel of Figure 15
which shows no saturation phenomenon in the dependence
of the characteristic temperatures on $\Upsilon$ for the helix.
Instead, both
$T_{min}$ and $T_f$  grow with $\Upsilon$ except at
small values of the parameter. However, the folding times and
the distributions of the defects in chirality do saturate with
$\Upsilon$ around the value of 1 so from now on we consider
the canonical case of $\Upsilon =1$.\\

Figure 16 compares the distribution and localization of the
defects in chirality obtained by incorporating $V^{CHIR}$ to those with
$V^{ANG}$. In both cases, the folding is declared by means of the Q criterion.
It is seen that the angular terms act about the same,
or slightly better, at
eliminating the defects in the helix 
(the defects with $V^{ANG}$ are also found to be delocalized) 
but somewhat worse
for the hairpin and crambin (not shown). 
In crambin, $V^{CHIR}$ my yield up to 14 defects whereas $V^{ANG}$
up to 9. We find, however, 
that the computer time  is significantly
shorter when using $V^{CHIR}$ 
than $V^{ANG}$ because of the difference in the
effort taken to calculate the potentials and to derive forces.
Thus $V^{CHIR}$ appears to generate about the same 
conformational effect as $V^{ANG}$ but the calculations proceed faster.
A purist's approach would be to include both terms in the Hamiltonian.\\

\vspace*{0.5cm}

\section{The side groups}
\vspace*{0.5cm}

The origin of chirality effects in proteins sits in the atomic structure
of amino acids. Suppose we generalize our Go-like model so that it
includes the $C^{\beta}$ atoms in addition to the $C^{\alpha}$ of the
backbone. Our version of this generalization will be described in a
separate publication and it involves interactions 
between two kinds of effective
atoms representing particular amino acids. The extra degrees of freedom
generate a more sophisticated set of steric constraints.
Will these constraints be sufficient to account for the chiral
effects?\\

Figure 17 suggests that such a modelling of the side groups in itself
does not guarantee emergence of the correct chirality 
when the native contacts are established and, in fact,
fares worse than the simple Go-like model with the chirality term.
The side group modelling absed on the $C^{\beta}$ atoms still requires
to be augmented by the chirality potential. We note that the chirality
defects in the side group model are found to be delocalized.\\

We conclude that the chirality potential with $\kappa$ equal to 1 or larger
is a useful and important ingredient of Go-like models of proteins.
The chirality-related criteria of folding should also be
of value when considering models that go beyond the Go approximation.
Our results were illustrated for two specific examples of secondary
structures and one protein. However, similar results were found
for several other helices [ 1ifv(127-142) and 1f63(3-18)] another
1pga hairpin and two other proteins: 1rpo and 1efn. We have found
that the $\alpha$-protein 1rpo behaves like a helix in that it
responds correctly to the chirality potential better than
to the angular potential.\\

{\bf Acknowledgments}

We acknowledge support
from KBN in Poland  (grant number 2 P03B 025 13) and thank A.
Kolinski for letting us use his computer cluster.

\vspace*{1cm}


\newpage
%
%

\begin{figure*}
\epsfxsize=3.8in
\centerline{\epsffile{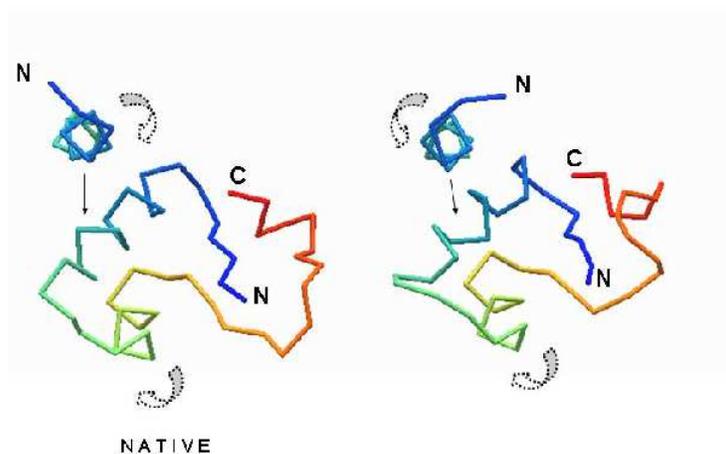}}
\caption{ 
Conformations of crambin in the backbone representation 
(there are 46 residues in crambin).
The figure on the left-hand side shows
the native structure (1crn in the Protein Data Bank).
The figure on the right-hand side corresponds to the conformation
in which all native contacts are established through a folding process
accomplished through a molecular dynamics evolution
in the Go-like model that is used in this paper. Even though
all amino acids are within the contact establishing distances
from each other in the folded state, some local chiralities are
opposite to what they should be.
The sense of chirality is shown by the spinning arrows.
The helical structure placed at the top of the figure winds
in opposite senses in the two panels.
}
\label{fig:crn}
\end{figure*}

\begin{figure*}
\epsfxsize=4.2in
\centerline{\epsffile{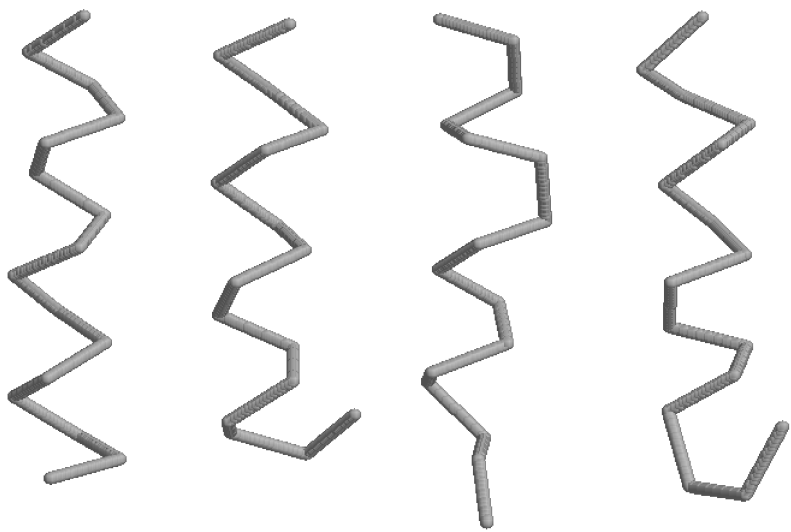}}
\caption{Backbone conformations of the $\alpha$-helix 1e6j(16-31),
extracted from the protein 1e6j.
The first conformation shown is native.
The remaining conformations are shown at a stage when folding is
considered to be accomplished. The time evolution is performed as the
molecular dynamics process in the Go-like model. The starting 
conformation corresponds to a straight line. The folding criteria are
Q, R, and A, left to right respectively.
The values of the local chiralities for $i$ running from
the second to the second from the end terminal amino acid are as follows.
As found in the true native state: 
0.83, 0.71, 0.75, 0.78, 0.81, 0.84, 0.80, 0.71, 0.81, 0.64, 0.84, 0.36,
0.63.\\
At folding declared by criterion Q:$\;\;$
0.84, 0.64, 0.82, 0.76, 0.71, 0.81, 0.64, 0.76, 0.74, 0.79, 0.46, 0.60,
-0.50.\\
At folding declared by criterion R:$\;\;$
-0.79, -0.77, -0.55, -0.77, -0.48, 0.66, 0.60, 0.79, 0.79, 0.59, 0.91, 
0.37, 0.34.\\
At folding declared by criterion A:$\;\;$
0.88, 0.45, 0.75, 0.69, 0.86, 0.87, 0.91, 0.66, 0.84, 0.66, 0.86, 0.39,
-0.18.}
\end{figure*}

\clearpage
\begin{figure*}
\epsfxsize=3.2in
\centerline{\epsffile{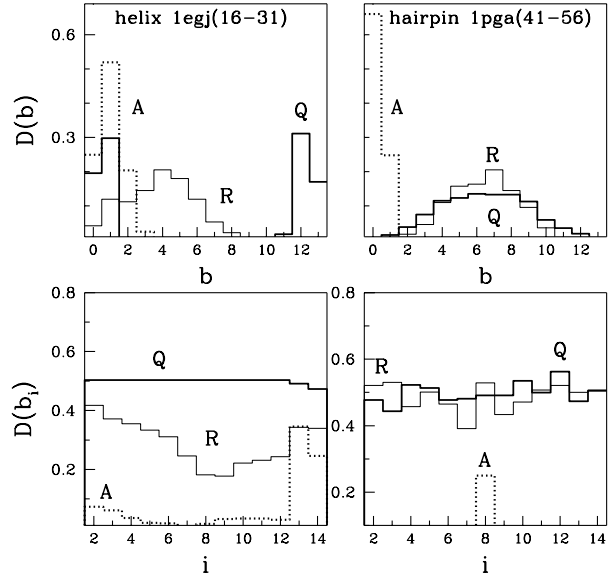}}
\caption{The top panels show distributions, $D(b)$, of the numbers, $b$,
of the wrong signed local chiralities when folding is declared according
to the criteria Q, A, and R, as indicated. The left and right panels are
for the helix 1e6j(16-31) and the hairpin 1pga(41-56) respectively.
The bottom panels show the sequential distribution of such chirality
defects.}
\end{figure*}

\begin{figure*}
\epsfxsize=3.2in
\centerline{\epsffile{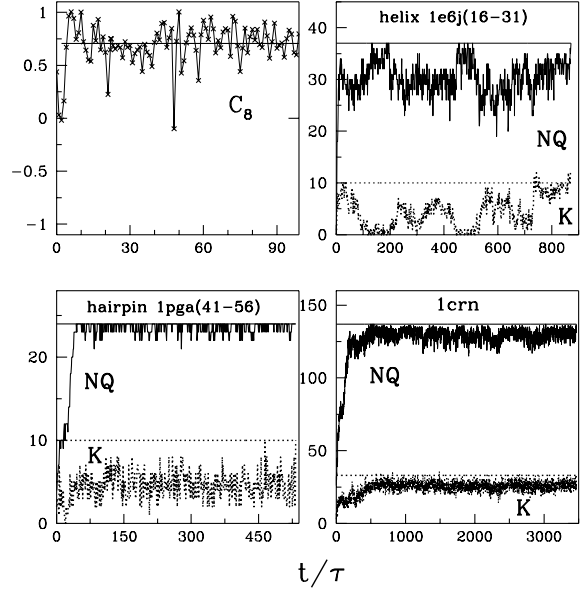}}
\caption{
The top left panel illustrates the time dependence of the local 
chirality for a single amino acid -- the central amino acid in the
1pga(41-56) hairpin. 
The remaining panels show examples of the time dependence
of the number of established native contacts, $NQ$, and the number
of established correct chiralities, $K$, in single trajectories
for the systems indicated. The evolution is stopped when $Q$ becomes 
equal to 1 and $K$ achieves the 75\% level of the native value.
These target thresholds are indicated by the horizontal lines.
 }
\end{figure*}

\begin{figure*}
\epsfxsize=4.0in
\centerline{\epsffile{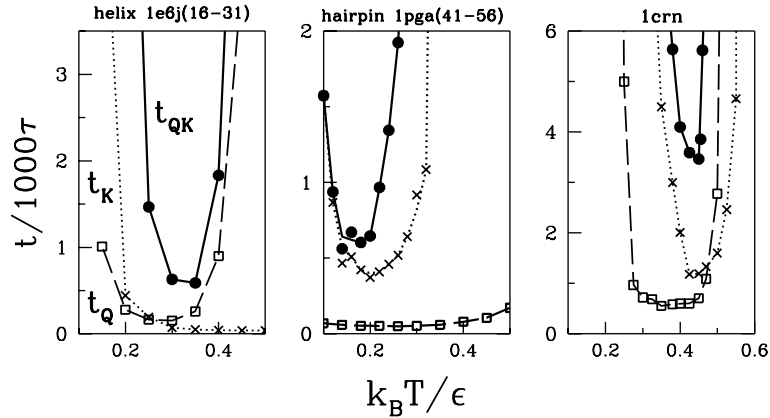}}
\vspace*{-0.4cm}
\caption{ 
The median folding times of the systems indicated at the top of the panels.
The data are based on at least 101 trajectories.
The dashed lines and the square data points correspond to the contact
based folding criterion Q. The solid line and the circle data points
correspond to the combined contact and chirality based criterion QK.
The dotted lines and the crosses correspond to the times needed to
establish 75\% of the local chiralities in the native fashion.
}
\end{figure*}

\begin{figure*}
\epsfxsize=4.0in
\centerline{\epsffile{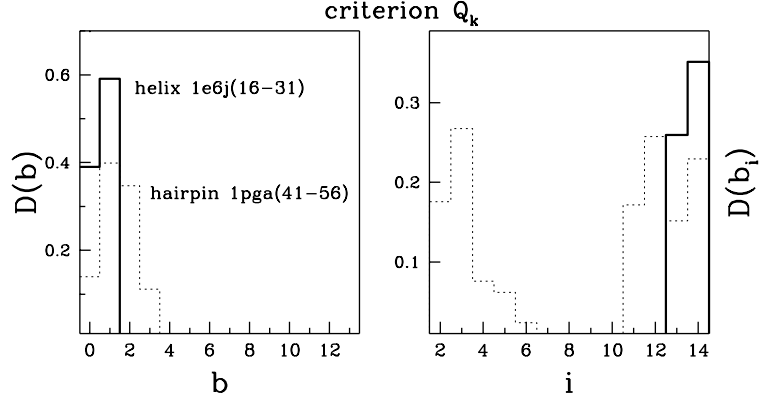}}
\vspace*{-0.4cm}
\caption{ 
The nature of chirality defects in conformations found through
the compound criterion Q$_K$ for the helix (the solid lines)
and the hairpin (the dashed lines). The left hand panel shows the
distribution of the sizes of the defects whereas the right hand panel
the distribution of their sequential location.
}
\end{figure*}

\begin{figure*}
\epsfxsize=3.2in
\centerline{\epsffile{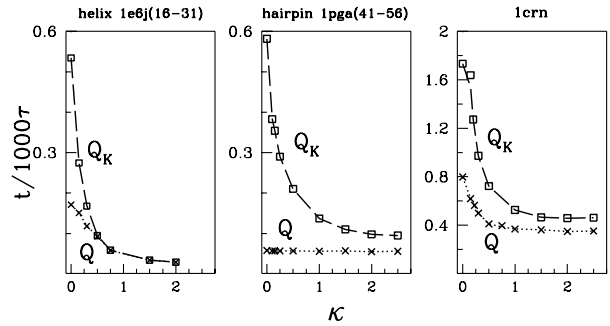}}
\caption{ 
Median folding times for the systems indicated at the top of each
panel as a function of the strength of the chirality potential.
The square data symbols correspond to the Q$_K$ criterion
and the crosses to the Q criterion.
The data points were obtained at $T_{min}$ which depends
on $\kappa$ and the criterion.
}
\end{figure*}

\begin{figure*}
\epsfxsize=4.0in
\centerline{\epsffile{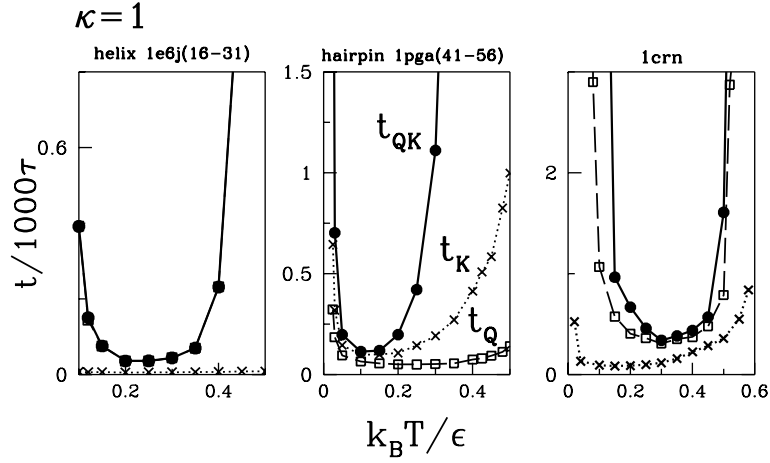}}
\vspace*{-0.3cm}
\caption{ 
Same as in Figure 5 but with the chiral potential
added to the Hamiltonian of the system. For the helix data shown in the
left panel, $t_Q$ and $t_{KQ}$ coincide within the scale of the figure.
}
\end{figure*}

\begin{figure*}
\epsfxsize=3.2in
\centerline{\epsffile{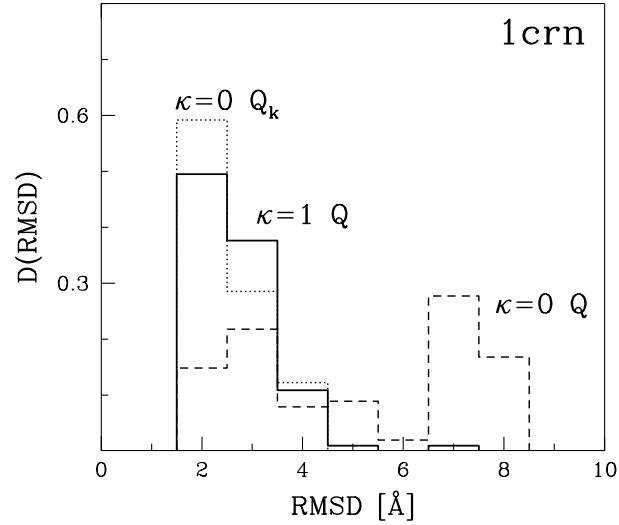}}
\vspace*{-0.3cm}
\caption{ 
Distribution of the RMSD distances away from the native state at folding
declared by the Q criterion. The dashed line corresponds to the
Hamiltonian with no chirality potential and the solid line to the
with the chirality term with $\kappa$=1. The data are collected at the
temperature of optimal folding and are based on 101 trajectories.
}
\end{figure*}

\vspace*{-1cm}

\begin{figure*}
\epsfxsize=3.8in
\centerline{\epsffile{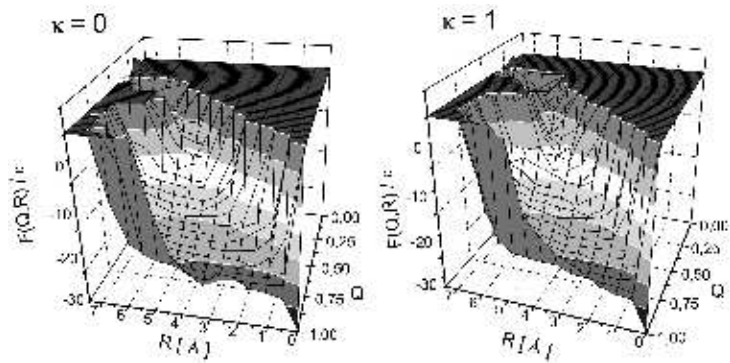}}
\caption{ 
The free energy corresponding to specific values of $Q$ and $R$ (the
RMSD distance away from the native state) for the helix at 
$\tilde{T}=0.3$. The figure on the right hand
side is for the system with the chirality potential with $\kappa$=1.
The figure on the left hand side involves no chirality potential.
The results are based on 100 trajectories of 20000 $\tau$ that start
from unfolded conformations.
}
\end{figure*}

\begin{figure*}
\epsfxsize=4.0in
\centerline{\epsffile{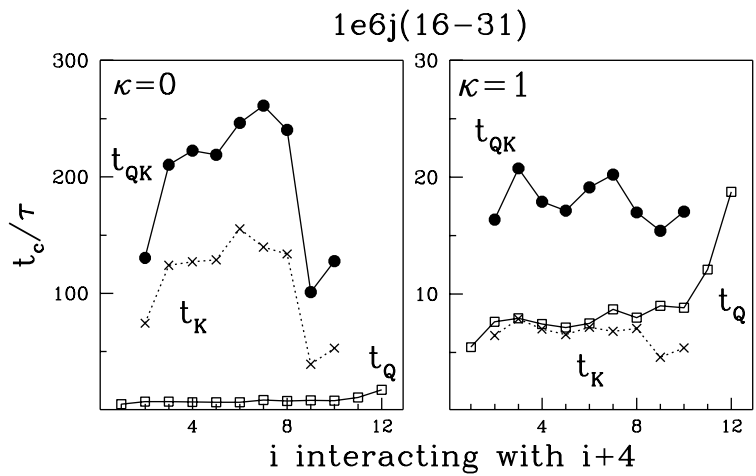}}
\vspace*{-0.3cm}
\caption{
The folding scenario for the $\alpha$ helix 1e6j(16-31) with (the right
panel) and without (the left panel) the chirality potential.
The building of the structure is illustrated through the contacts
that correspond to the hydrogen bonds -- between amino acid $i$ and
$i+4$.
$t_Q$ denotes the average first time to establish
the contact between $i$ and $j$
whereas $t_K$ is the average first time at which the
local chiralities of both amino acids are native like.
$t_{QK}$ denotes the average first time at which both the contact and the
chiralities are set correctly. The data points are averaged over 2000
trajectories.
}
\end{figure*}

\begin{figure*}
\epsfxsize=4.0in
\centerline{\epsffile{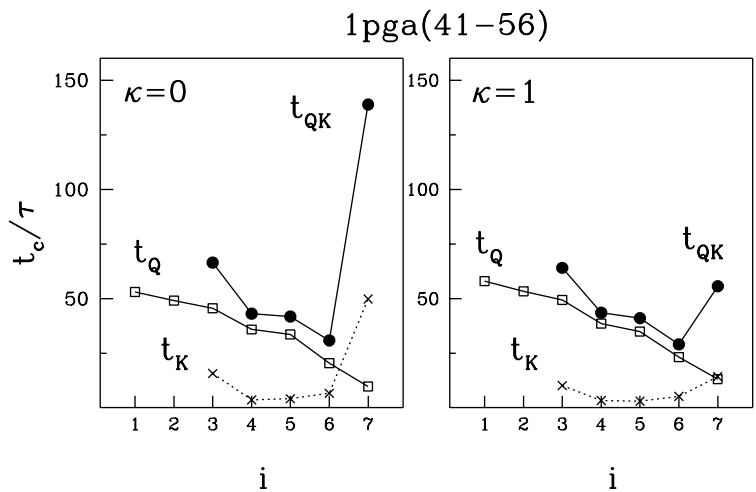}}
\caption{Similar to figure 11 but for the  hairpin 1pga(41-56).
The focus here is on the hydrogen bonds between $i$ and $N-i+1$.
}
\end{figure*}

\begin{figure*}
\epsfxsize=4.0in
\centerline{\epsffile{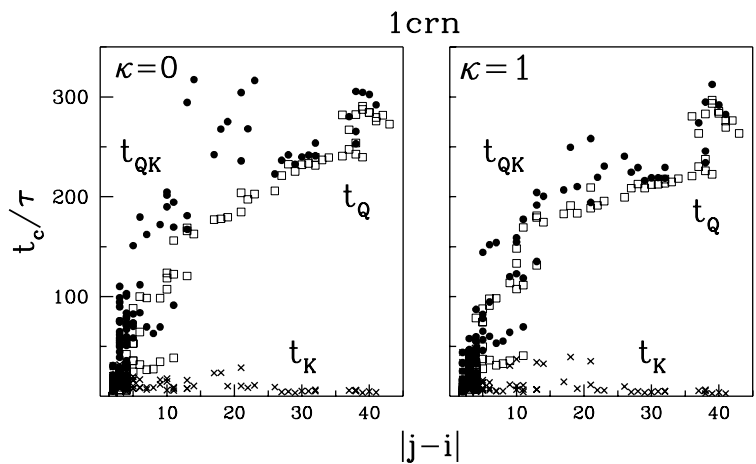}}
\caption{
The folding scenario of crambin with (the right panel)
and without (the left panel) the chirality potential. $|j-i|$
denotes the sequential distance between two amino acids, $j$ and $i$.
The symbols used are as in figure 11.
}
\end{figure*}

\begin{figure*}
\epsfxsize=4.0in
\centerline{\epsffile{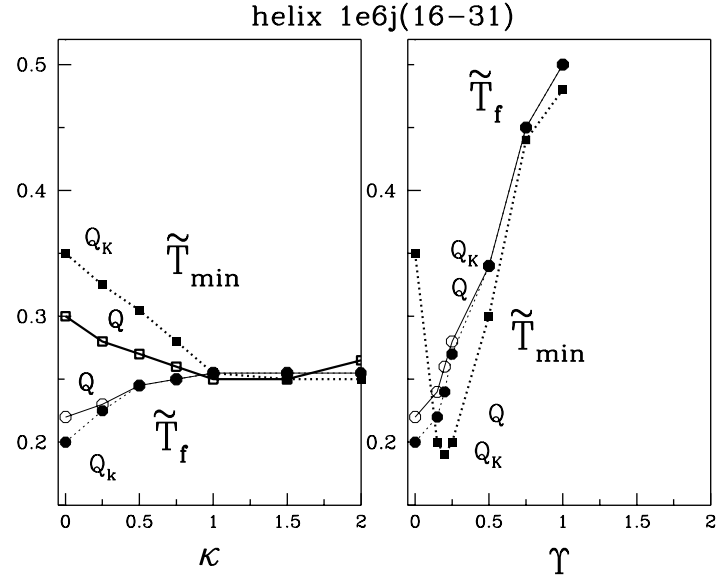}}
\caption{The left panel shows the dependence of 
$\tilde{T}_{min}$ and $T_f$ on the strength of the chirality
potential. Q and Q$_K$ denote the methods used to define
the native basin. The right panel shows similar quantities
but as a function of the amplitude of the angle-dependent
potential. 
}
\end{figure*}

\begin{figure*}
\epsfxsize=4.0in
\centerline{\epsffile{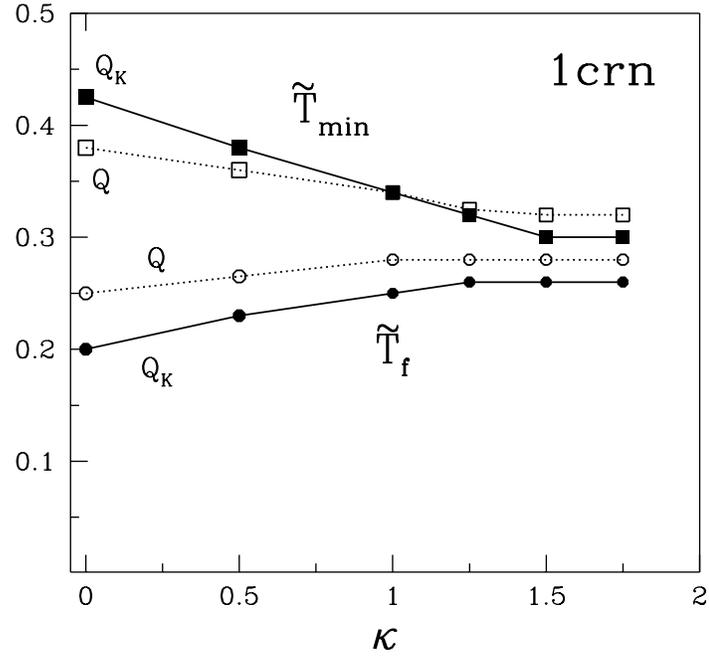}}
\caption{
Similar to the left hand panel of Figure 14 but for crambin.
}
\end{figure*}

\begin{figure*}
\epsfxsize=4.0in
\centerline{\epsffile{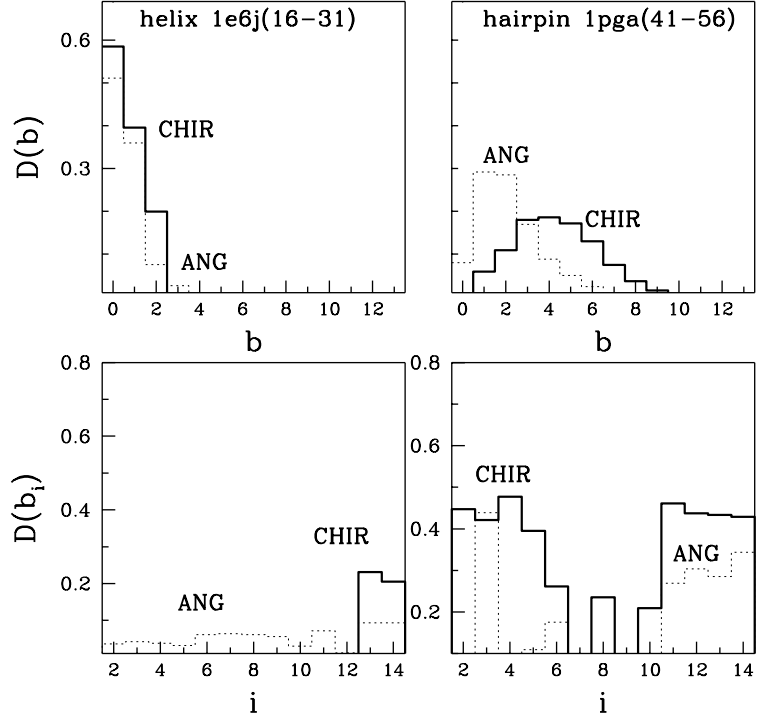}}
\caption{Same as in Figure 3 but with the chiral potential
( solid line) or with the angular potential  (dotted line)
added to the contact interactions and the
tethering terms. Folding is decided based on the Q criterion.
}
\end{figure*}

\begin{figure*}
\epsfxsize=4.0in
\centerline{\epsffile{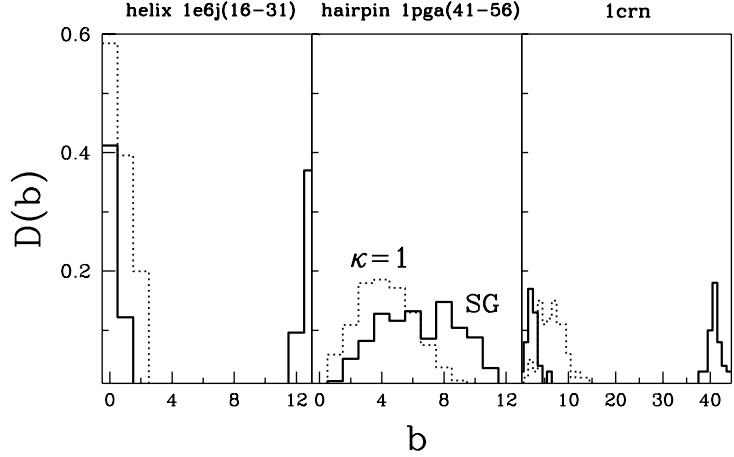}}
\caption{Distribution of the wrong signed chiralities for the systems
indicated and modelled by the Go-like Hamiltonian with the side groups 
(SG - the solid lines) and by the Go-like Hamiltonian without
the side groups but with
the chirality potential corresponding to $\kappa$=1 (the dotted line).
The Q criterion is used here to determine folding.
}
\end{figure*}

\end{document}